\documentclass[aps,prb,twocolumn,amsmath,amssymb,superscriptaddress,floatfix]{revtex4-2}

\usepackage{sidecap}

\usepackage{amsmath}
\usepackage{xcolor}
\usepackage{graphicx}
\usepackage{dcolumn}
\usepackage{bm}
\usepackage{verse}
\usepackage[]{graphicx}      
\usepackage{bm}             	
\usepackage{times}			
\usepackage{tabularx}
\usepackage{color}
\usepackage{dcolumn}		
\usepackage{amsmath}
\usepackage{amssymb}
\usepackage{amsfonts}
\usepackage{times}
\usepackage{tabularx}
\usepackage{xcolor,colortbl}
\usepackage{rotating}

\usepackage{amsmath}

\makeatletter
\newcommand\xleftrightarrow[2][]{%
  \ext@arrow 9999{\longleftrightarrowfill@}{#1}{#2}}
\newcommand\longleftrightarrowfill@{%
  \arrowfill@\leftarrow\relbar\rightarrow}
\makeatother

\begin{document}
\title{Measuring a population of spin waves from the electrical noise of an inductively coupled antenna}
\author{T. Devolder}
\affiliation{Universit\'e Paris-Saclay, CNRS, Centre de Nanosciences et de Nanotechnologies, 91120, Palaiseau, France}
\email{thibaut.devolder@u-psud.fr}
\author{S.-M. Ngom}
\author {A. Mouhoub}
\author {J. L\'etang}
\author{J.-V. Kim}
\author{P. Crozat}

\author{J.-P. Adam}
\affiliation{Universit\'e Paris-Saclay, CNRS, Centre de Nanosciences et de Nanotechnologies, 91120, Palaiseau, France}
\author{A. Solignac}
\affiliation{SPEC, CEA, CNRS, Universit\'e Paris-Saclay, 91191 Gif-sur-Yvette, France}
\author{C. Chappert}
\affiliation{Universit\'e Paris-Saclay, CNRS, Centre de Nanosciences et de Nanotechnologies, 91120, Palaiseau, France}
\date{\today}                                           
                       
%
%
\begin{abstract}
We study how a population of spin waves can be characterized from the analysis of the electrical microwave noise delivered by an inductive antenna placed in its vicinity.  
The measurements are conducted on a synthetic antiferromagnetic thin stripe covered by a micron-sized antenna that feeds a spectrum analyser after amplification. The antenna noise contains two contributions. The population of incoherent spin waves generates a fluctuating field that is sensed by the antenna: this is the "magnon noise". The antenna noise also contains the contribution of the electronic fluctuations: the Johnson-Nyquist noise. The latter depends on all impedances within the measurement circuit, which includes the antenna self-inductance. As a result, the electronic noise contains information about the magnetic susceptibility of the stripe, though it does not inform on the absolute amplitude of the magnetic fluctuations. 
For micrometer-sized systems at thermal equilibrium, the electronic noise dominates and the pure magnon noise cannot be determined. If in contrast the spinwave bath is not at thermal equilibrium with the measurement circuit, and if the spin wave population can be changed then one could measure a mode-resolved effective magnon temperature provided specific precautions are implemented. 

\end{abstract}

\maketitle

%
%

\section{Introduction}

Spin waves (SW) possess specific properties that make them well-suited for microwave applications \cite{Kruglyak_magnonics_2010, chumak_magnon_2015}. This includes anisotropic and non-reciprocal dispersion relations as well as tunability. Moreover, there is a growing use of the non-linear interactions of spin waves for advanced computing proposals \cite{papp_nanoscale_2020, papp_characterization_2021, wang_nonlinear_2020}; these applications rely on magnetic bodies where large populations of coherently pumped spin waves share a common space with their less coherent biproducts \cite{naletov_ferromagnetic_2007, bauer_nonlinear_2015} and the thermal populations of spin waves \cite{demidov_nonlinear_2011, schultheiss_excitation_2019, kamimaki_parametric_2020}. It is thus of interest to develop experimental techniques able of measuring spin waves in broad frequency intervals with a large dynamic range, ideally from the floor of the thermal population of spin waves up to the regime of large amplitudes of magnetization precession. There is also a fundamental interest in the development of such experimental techniques since observing the fluctuations of a system is a way \cite{raquet_electronic_2001} to get direct insight into its intrinsic internal dynamics.

Brillouin Light Scattering (BLS) \cite{hillebrands_progress_1999} and its space-resolved variant \cite{demokritov_micro-brillouin_2008, sebastian_micro-focused_2015} are the techniques of choice for SW measurements: they feature the proper sensitivity and dynamic range, but they involve a somewhat heavy infrastructure for a lateral resolution that is limited by diffraction. A better lateral resolution --essentially limited by nanofabrication \cite{ciubotaru_all_2016}-- is provided by the electrical methods such as propagating spin wave spectroscopy (PSWS \cite{bailleul_propagating_2003, devolder_measuring_2021}) or cantilever-based techniques like magnetic resonance force microscopy \cite{klein_ferromagnetic_2008}. Unfortunately most of these techniques rely on the homodyne detection of coherent SW that must be pumped in an ad-hoc manner, with no ability to measure spin waves out-of-sync from the pump. 
Finally, broadband magneto-resistive noise spectroscopy is able to measure SW thermal populations even in sub-100 nm magnet \cite{helmer_quantized_2010}, however it requires to apply invasive currents and can be implemented only on patterned systems that display magneto-resistance. A versatile electrical method capable of measuring SW populations over a large dynamic range in any material is therefore still missing. 

Here we study whether the noise of a lithographically-defined inductive antenna can be used to measure a population of incoherent SW placed in its vicinity. We illustrate the method on a synthetic antiferromagnet (SAF) in which both the SW frequencies and the SW group velocities can be adjusted conveniently over large intervals using external fields. \textcolor{black}{This method can actually be implemented on any type of magnet, not just SAFs}. We show that the method is versatile, broadband but requires the implementation of impedance matching precautions. 

Our samples are described in section \ref{samples}. Section \ref{setup} depicts two experimental configurations, their operation, and provides the results on the electrical power spectra delivered by the antenna. The results indicate that the magnetization-sensitive part of the signal delivered by the antenna arises from two contributions. We conjecture that the two origins are uncorrelated so that their noise powers are additive. Section \ref{Johnson-Nyquist} evaluates how the field- and frequency-dependent impedance of the antenna modulates the overall electronic noise power that it delivers, independently from the actual population of spin waves that would be present in the sample in the absence of the antenna. Section \ref{sectionEmf} accounts for the power collected by the inductive antenna thanks to the magnetic fluctuations underneath, assumed to be related to the population of incoherent SW. We then compare the amplitudes of these two sources of signal. We finally discuss the conditions on the magnetic material, geometry and circuit impedance that must be met in order to be able to measure quantitatively a thermal or non-thermal population of incoherent spin waves from the total noise delivered by the antenna.

\section{Samples} \label{samples}
\subsection{Materials, geometry and uniform resonances}
\begin{figure}
\hspace*{-0.2cm}\includegraphics[width=7.8 cm]{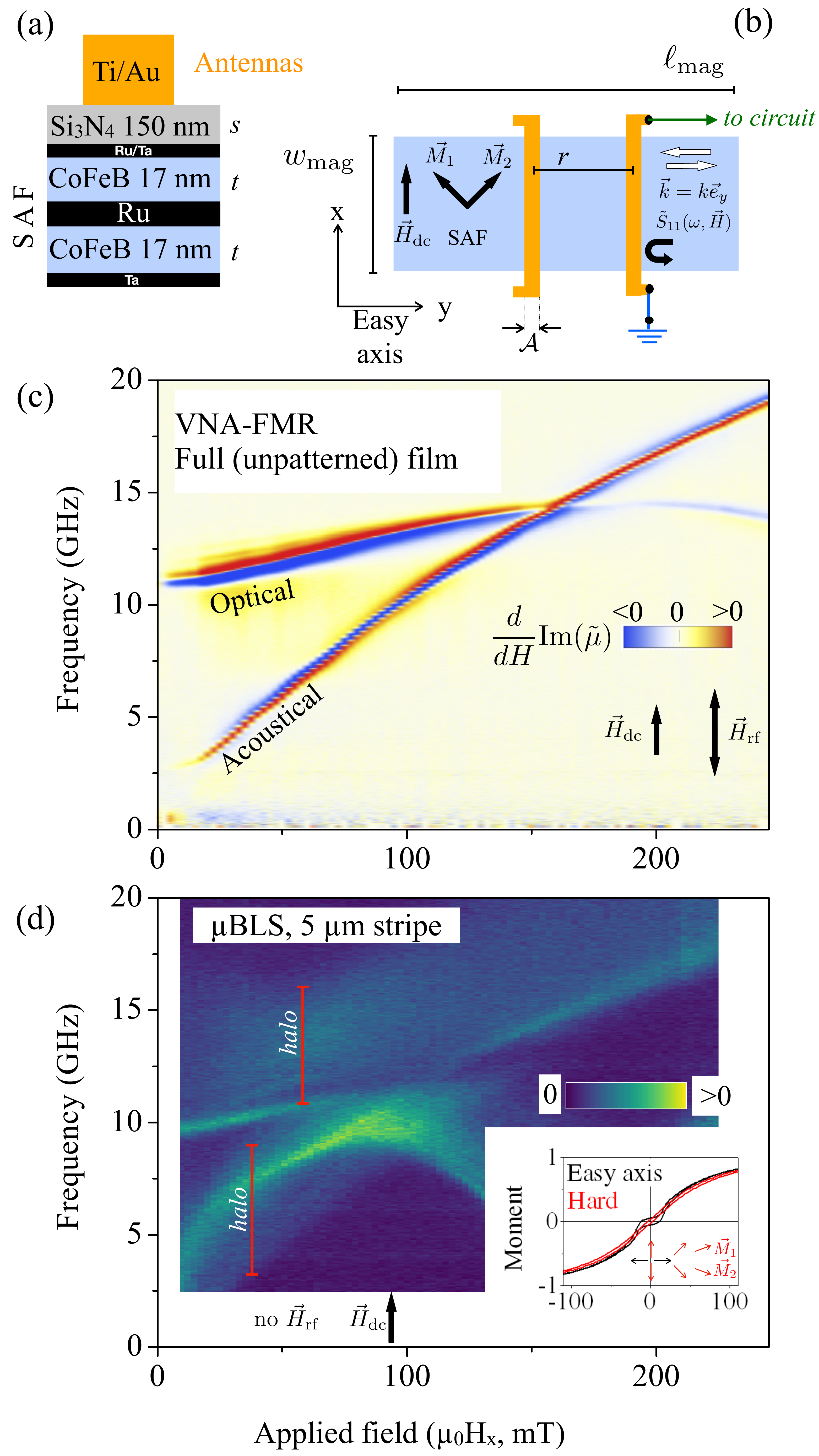}{\centering}
\caption{Properties of Synthetic Antiferromagnet (SAF) films. (a): Sketch of the stack. (b):  Sketch of the device and main notations. (c): Field derivative of the imaginary part of the permeability of the film measured by VNA-FMR prior to patterning \textcolor{black}{using an external stripline of width 0.05 mm and an rf power of 0 dBm}. The optical and acoustical spin wave resonances at zero wavevector have frequencies corresponding to the red/blue frontiers. (d): Spin wave thermal spectra measured by BLS microscopy at the middle of a 5 $\mu$m wide SAF stripe, on a film with thinner (hence more transparent) Ta cap. The red vertical bars illustrate the frequency range in which thermal spin waves are detected. Inset: hysteresis loops along the easy and hard axes.}
\label{FigureFILM-FMR-BLS}
\end{figure}

We use in-plane magnetized Synthetic AntiFerromagnetic (SAF) films of composition Ta (6 nm) / Co$_{40}$Fe$_{40}$B$_{20}$ ($t$) /Ru (7 \r{A}) / Co$_{40}$Fe$_{40}$B$_{20}$ (t) / Ru (0.4 nm)/ Ta (6 nm) with thicknesses $t=17~\textrm{nm}$ [Fig.~\ref{FigureFILM-FMR-BLS}(a)]. The magnetization and the damping of the CoFeB material were determined on single films to be respectively $M_s=1.35~\textrm{MA/m}$ and $\alpha\approx 0.004$. A small uniaxial anisotropy field of 3 mT exists within the sample plane; the spin-flop field of the SAF is thus finite ($\approx$0-15 mT) \cite{worledge_spin_2004, devolder_spin_2012} except when the field is applied exactly along the hard axis. In this paper, the hard axis is taken conventionally along $x$ [Fig.~\ref{FigureFILM-FMR-BLS}(a)], although the anisotropy field will be neglected in the equations.

%

Vector Network Analyser ferromagnetic resonance (VNA-FMR~\cite{bilzer_vector_2007}) was used to check the properties of the complete SAF stack on an unpatterned film [Fig.~\ref{FigureFILM-FMR-BLS}(c)]. The parallel pumping configuration was chosen to allow detecting both the acoustical and the optical eigenmodes \cite{zhang_angular_1994}, all at zero wavevector $k$. At remanence, the uniform (i.e. $k=0$) optical mode is at $\frac{\omega_\textrm{op}}{2 \pi} = 11.3~\textrm{GHz}$. The interlayer exchange field $\mu_0 H_{j}=-\frac{2 J}{ {M_{s}}{t}}$ (with $J$ the interlayer exchange energy) can be estimated from the 2-macrospin approximation \cite{devolder_spin_2012}. The formulas are detailed in Table I. A first estimate stems from the optical frequency $\omega_\textrm{op} \big|_{H_x=0}=\gamma_0 \sqrt{M_s H_j }$; it yields $J \approx-1~\textrm{mJ/m}^2$. An alternative estimate can be deduced from a linear fit of the frequency of the uniform (i.e. $k=0$) acoustical mode $\frac{\omega_\textrm{ac}}{2 \pi}$ versus $H_x$ at low field (see appendix I). This alternative estimation yields a slightly higher value of $J \approx-1.2~\textrm{mJ/m}^2$. We noticed that this value is very sensitive to the details of material deposition, thermal treatment and aging.

For this study the SAF films were patterned into stripes of width $w_\textrm{mag} = 20 ~\mu \textrm{m}$ along $(x)$ and lengths $\ell_\textrm{mag}=75~\mu\textrm{m}$ along $(y)$ [Fig.~\ref{FigureFILM-FMR-BLS}(b)]. We then deposit a $s=150$ nm insulation layer of Si$_3$N$_4$. We finally make single-wire antennas of composition Ti(10 nm)/Au(150 nm), width $\mathcal{A}=1.8~\mu \textrm{m}$. The antenna \textit{dc} impedance amounts to $\tilde{Z}_\textrm{out} \approx(10 + 0 i)\Omega$. \textcolor{black}{. Although our present purpose --the inductive measurement of the spin wave populations-- requires only one antenna we have deposited a second one at a distance $r=5~\mu \textrm{m}$ to enable to check of the SW main properties using propagating spin wave spectroscopy (not shown). }
%
\begin{figure*}
\hspace*{-0.1cm}\includegraphics[width=14 cm]{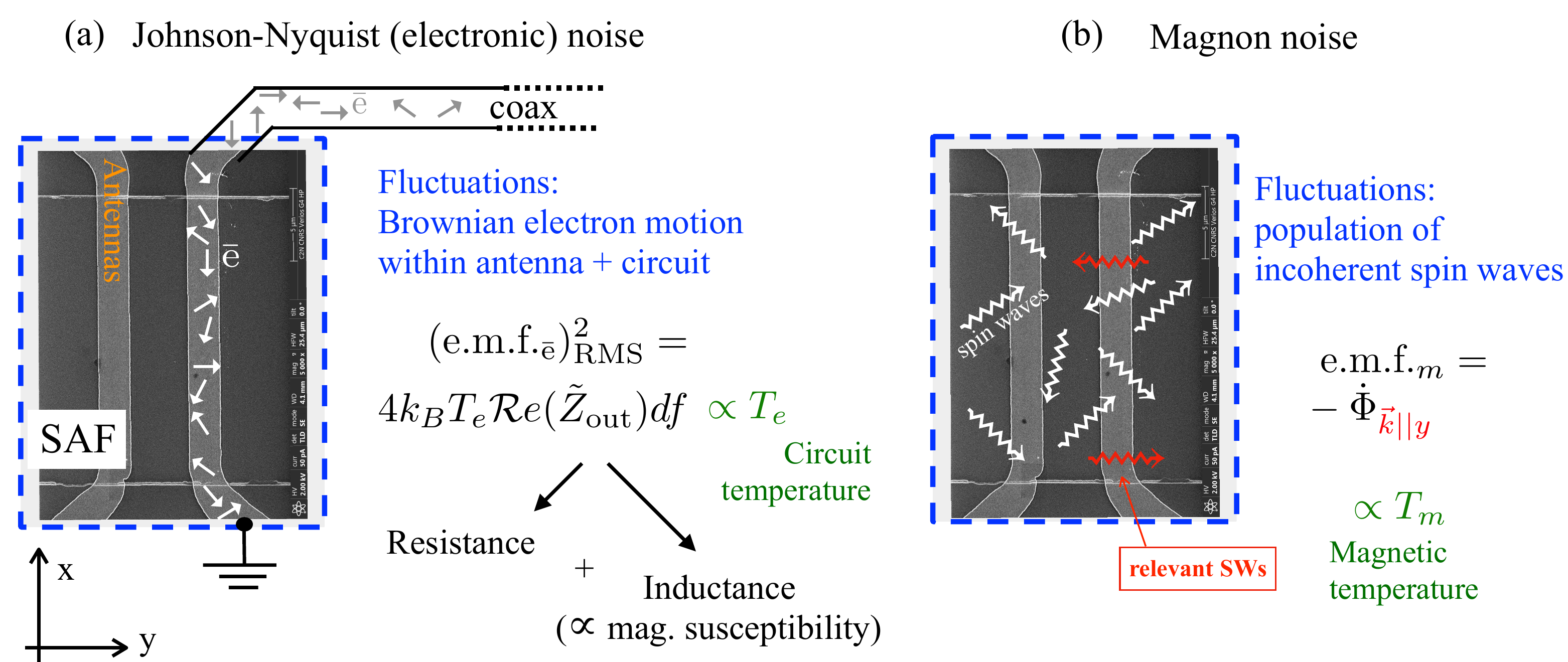}{\centering}
\caption{Noise sources within the sample: (a) Variance of the Johnson-Nyquist e.m.f due to the Brownian motion of the electrons within the measurement conductive and dissipative path, and (b) inductive e.m.f. arising from the fluctuations within the spin wave bath underneath the antenna. }
\label{figEnergySources}
\end{figure*}

\subsection{Identification of spin wave branches using microfocused BLS} 

To characterize the spin waves in our material, we measured their thermal spectra by BLS with a laser focused at the diffraction limit (350 nm) at the surface of a SAF [Fig.~\ref{FigureFILM-FMR-BLS}(d)]. We used a dedicated sample in which the thickness of the Ta cap layer was reduced to 5 nm to allow optical access to the topmost CoFeB layer. Besides, the stripe width was 5 $\mu$m.
In this focused optical configuration, the sensitivity is maximal at ${k}={0}$. The similarity between the BLS and the VNA-FMR spectra [Fig.~\ref{FigureFILM-FMR-BLS}(c) vs (d)] confirms the identification of the modes even if the mode frequencies of the BLS-dedicated sample and the fully capped samples are slightly different. We believe that this difference is due to the large sensitivity of the interlayer exchange coupling to the structural properties of the material.

Interestingly, this focused BLS measurement is also sensitive to the SW with finite wavenumbers, typically up to $k_\textrm{max}=18 ~\textrm{rad}/\mu$m. For spin waves with wavevector orientation $\theta = \{\vec{k}, \vec{H_x}\}$, the frequency band in which the SW can be optically detected has a width $\frac{1}{2 \pi}{v_{g}(\theta) k_\textrm{max}}$, where $v_g(\theta)=\frac{\partial \omega}{\partial |k|}$ is the (positive or negative) group velocity in the $\theta$ direction. The optical SW of a SAF are always \textit{forward} waves such that the thermal population of optical SW is seen as a halo extending only \textit{above} the frequency of the $k=0$ optical mode [Fig.~\ref{FigureFILM-FMR-BLS}(d)]. This contrasts with the acoustical SW for which the dispersion relation includes both \textit{forward and backward} waves as $\theta$ is varied, such that the halo extends both \textit{below and above} the frequency of the $k=0$ acoustical mode. 

We emphasize that in the remainder of this study we will use a linear antenna instead of the circular laser spot. As a result, we shall be sensitive only to spin waves with either vanishing wavevectors, or with a wavevector perpendicular to the antenna (the inductive contributions of other SW average out over the antenna length). In this $\vec k \perp \vec{H_x}$ configuration, both the optical and the acoustical SWs have a forward character (see appendix); their frequencies will always be above the $k=0$ cases. This is a major difference from the BLS case.

\section{Measurement of the electrical noise power delivered by the inductive antenna} \label{setup}
\subsection{Qualitative description of the sample impedance and fluctuators}%
Our goal is now to determine the different contributions to the rf electrical noise power that the antenna can deliver. For qualitative understanding, it is convenient to view the \{antenna + SAF\} system as the noisy electromotive force $\textrm{(e.m.f)}$ of a Thevenin generator connected in series with a noise-free output impedance $\tilde{Z}_\textrm{out}$ (Fig.~\ref{figEnergySources}). The impedance $\tilde{Z}_\textrm{out}$ comprises a (non-stochastic) resistive part and a (non-stochastic) self-inductance part which increases linearly with the magnetic susceptibility of the SAF. As a result, the impedance matching between the sample and the measurement circuit depends on the applied field and on the frequency.

Besides, the sample contains two a priori independent noise sources that contribute to its e.m.f.: the electronic fluctuations within the conductors [the "Johnson-Nyquist noise", Fig.~\ref{figEnergySources}(a)] and the magnetic fluctuations inductively collected by the antenna [the "magnon noise", Fig.~\ref{figEnergySources}(b)]. 
\subsection{Experimental configuration n°1: the conventional apparatus}%
\begin{figure*}
\hspace*{-0.1cm}\includegraphics[width=14 cm]{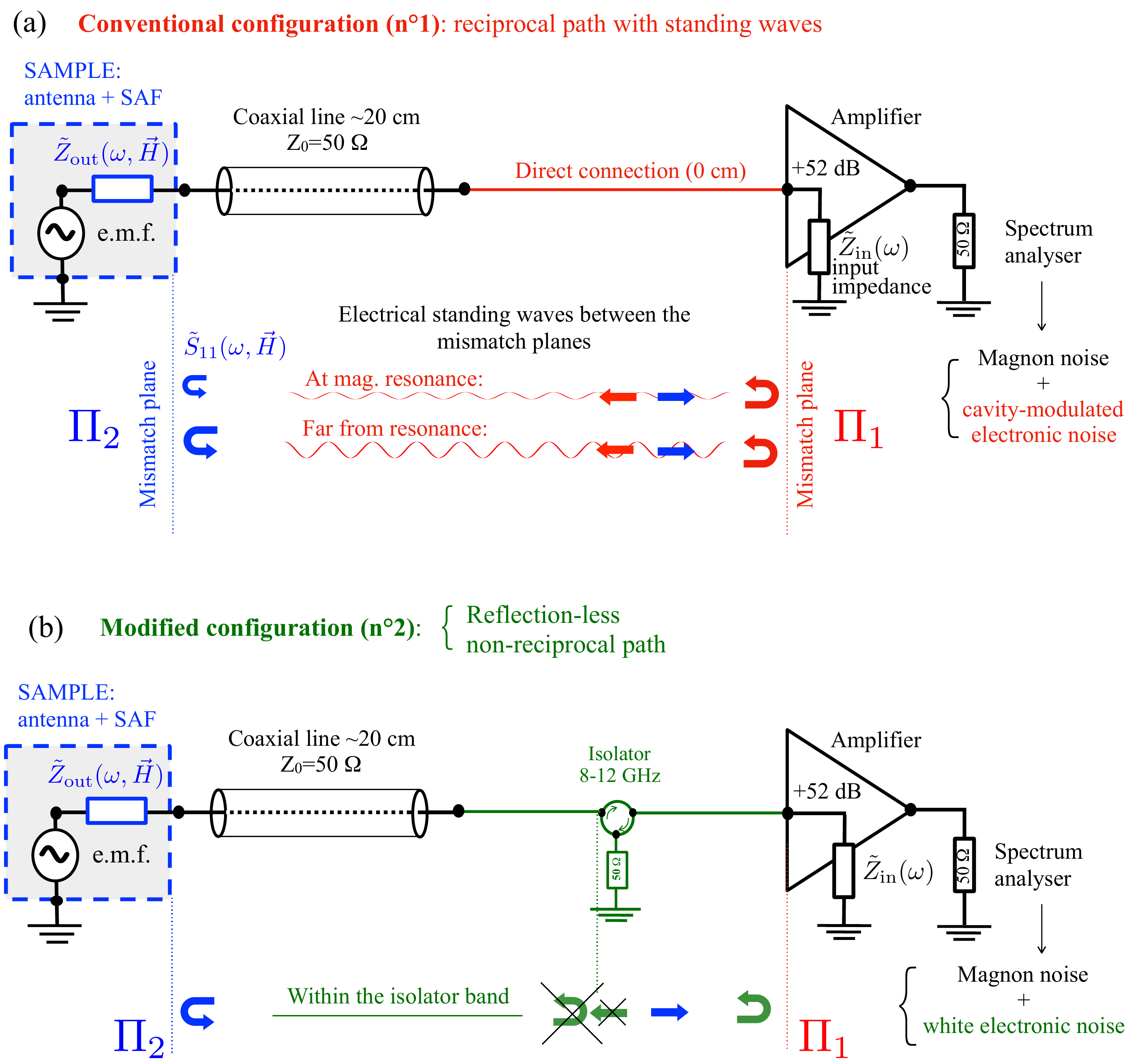}{\centering}
\caption{Experimental set-ups for the measurement of the spectral density of the electrical power delivered by the antenna. (a): The first "conventional" configuration (red color) allows the electrical signal to flow back (red arrows) and forth (blue arrows) between the antenna and the amplifier which both partially reflect at the planes $\Pi_1$ and $\Pi_2$, thereby forming a cavity hosting standing microwaves. (b) The second "modified" configuration (green color) includes an isolator that prevents the back flow (hatched green arrow) of electromagnetic energy from the amplifier towards the antenna. The standing waves are largely mitigated in the operating range of the isolator.}
\label{figSetUp}
\end{figure*}

In a first step, the set-up chosen to spectrally analyze this e.m.f is inspired from the ones widely used for the characterization of spin-torque oscillators \cite{rippard_direct-current_2004} or hard-disk read-heads \cite{pauselli_magnetic_2017}: the antenna feeds a spectrum analyzer (SA) using an using an \textit{rf} probe and a coaxial cable of characteristic impedance $Z_0=50~\Omega$, followed by  a broadband amplifier [cf. Fig.~\ref{figSetUp}(a)] . The amplifier gain ($G_\textrm{ampli}=52~\textrm{dB}$) is chosen large enough to bring the sample noise above the one internally generated at the front end of the SA (typically 24 dB above thermal noise). The amplifier noise figure is chosen as low as possible (here: 2 dB). 
%

Note that in broadband amplifier technologies, a compromise is done between minimal noise figure and perfect impedance matching: the amplifier input impedance is $\tilde{Z}_\textrm{in}(\omega) \neq Z_0$; this mismatch leads to a partial reflection at the amplifier's input (here: $ -17 \textrm{~dB} \leq S_{11}^\textrm{amp} \leq -10~\textrm{dB}$ in the used frequency band). As a result, there are inevitably \textit{two} mismatch planes [labeled $\Pi_1$ and $\Pi_2$ in Fig.~\ref{figSetUp}] in this conventional measurement circuit. The voltage/current electromagnetic waves can travel back and forth in the cavity formed between these two  planes: standing microwaves form within the coaxial cable. The interferences are constructive along a frequency comb of spacing $v_\varphi / \xi$, where $\xi$ is the length of the cable and $v_\varphi \approx 2.1 \times 10^8~\textrm{m/s}$ is the phase velocity of the electromagnetic waves in the coaxial cable. The voltage standing wave ratio depends on the reflection parameters at the two ends of the cavity; in particular, it depends on the applied field and on the frequency through $\tilde{Z}_\textrm{out}$, as illustrated in the inset of Fig.~\ref{FigureExpProcedureConfig1}(a) for two different applied fields.

\subsubsection{Measurement procedure}%
Our measurement procedure is illustrated in Fig.~\ref{FigureExpProcedureConfig1}. We first use a single port VNA to determine $\tilde{Z}_\textrm{out}(\omega, \vec{H})$ and $\tilde{Z}_\textrm{in}(\omega)$ separately. We then mount the measurement chain of Fig.~\ref{figSetUp}(a). The resolution bandwidth (RBW) of the spectrum analyser is set to its largest possible value (here $\textrm{RBW}=50~\textrm{MHz}$) to collect maximal energy and thereby speed up the acquisition. The magnitude of the noise power spectral density arriving at the SA is then typically of the order of the Johnson-Nyquist noise floor: $10~\textrm{Log}_{10}(k_B T_e \times \textrm{RBW}) + 30 + G_\textrm{ampli} \approx - 44~ \textrm{dBm}$ per investigated frequency point, where $T_e$ is the circuit temperature and the term '30' accounts for the mW to Watt conversion. This power spectral density [PSD, Fig.~\ref{FigureExpProcedureConfig1}(b)] displays a slow frequency dependence essentially set by the roll-off of $G_\textrm{ampli}(f) $, on top of which faster variations can be seen as a ripple. This ripple is a manifestation of the presence of standing electrical waves within the measurement circuit. 

We then record the variation of the PSD with the applied field $H_x$ and perform the field-to-field subtraction $\textrm{PSD} (H_x) - \textrm{PSD} (H_\textrm{ref})$ to reveal the tiny (typically a few 0.01 dB) variations of the PSD with the magnetic field, hereafter referred to as the 'excess noise' [Fig.~\ref{FigureExpProcedureConfig1}(c) and (d)]. $H_\textrm{ref}$ is a arbitrarily chosen reference field; here we have chosen $H_\textrm{ref}=0$. 
We strongly emphasize that subtracting $\textrm{PSD} (H_\textrm{ref})$ is just a \textit{convenient} way to reveal the dependence of the PSD with the applied field. As we will see later, this "(field-to-field) excess noise" resulting from this field-to-field subtraction is equivocal and has generaly no clear physical meaning. 
A quick look at Fig.~\ref{FigureExpProcedureConfig1}(d) indicates however immediately that the field-to-field excess noise when measured with this conventional apparatus contains magnetic information.

\subsubsection{Magnetic information within the field-to-field excess noise for the conventional experimental configuration}%
The field-to-field excess noise is found to be flat and featureless at the frequencies much below the uniform acoustical mode (10.3 GHz at 100 mT in Fig.~\ref{FigureExpProcedureConfig1}). However near and above that frequency, the excess noise becomes markedly oscillatory [Fig.~\ref{FigureExpProcedureConfig1}(c)]. Importantly, striking anticrossings are detected [Fig.~\ref{FigureExpProcedureConfig1}(d)] between the branch of uniform acoustical spin waves and the frequency comb of the $\Pi_1-\Pi_2$ microwave cavity. This is in stark contrast with the BLS result [Fig. 1(d)]. 
These anticrossings evidence the existence of a significant coupling occurring at $\Pi_2$ (i.e. $\tilde{Z}_\textrm{out}$) between the quasiparticles of the measurement system (the microwave photons confined in the $\Pi_1-\Pi_2$ cavity) and the magnons within the magnetic film. We wish to emphasize that these anticrossings were not present in the BLS measurement.

This leads us to question whether the conventional electrical experimental configuration actually measures the magnetic noise and/or whether it does so in a non-invasive manner. We have thus implemented the following modified measurement configuration meant to change the interaction between the spin wave bath and the measurement circuit.

\subsection{Experimental configuration n°2: \\for a reduced voltage standing wave ratio}%
\begin{figure}
\includegraphics[width=8.0 cm]{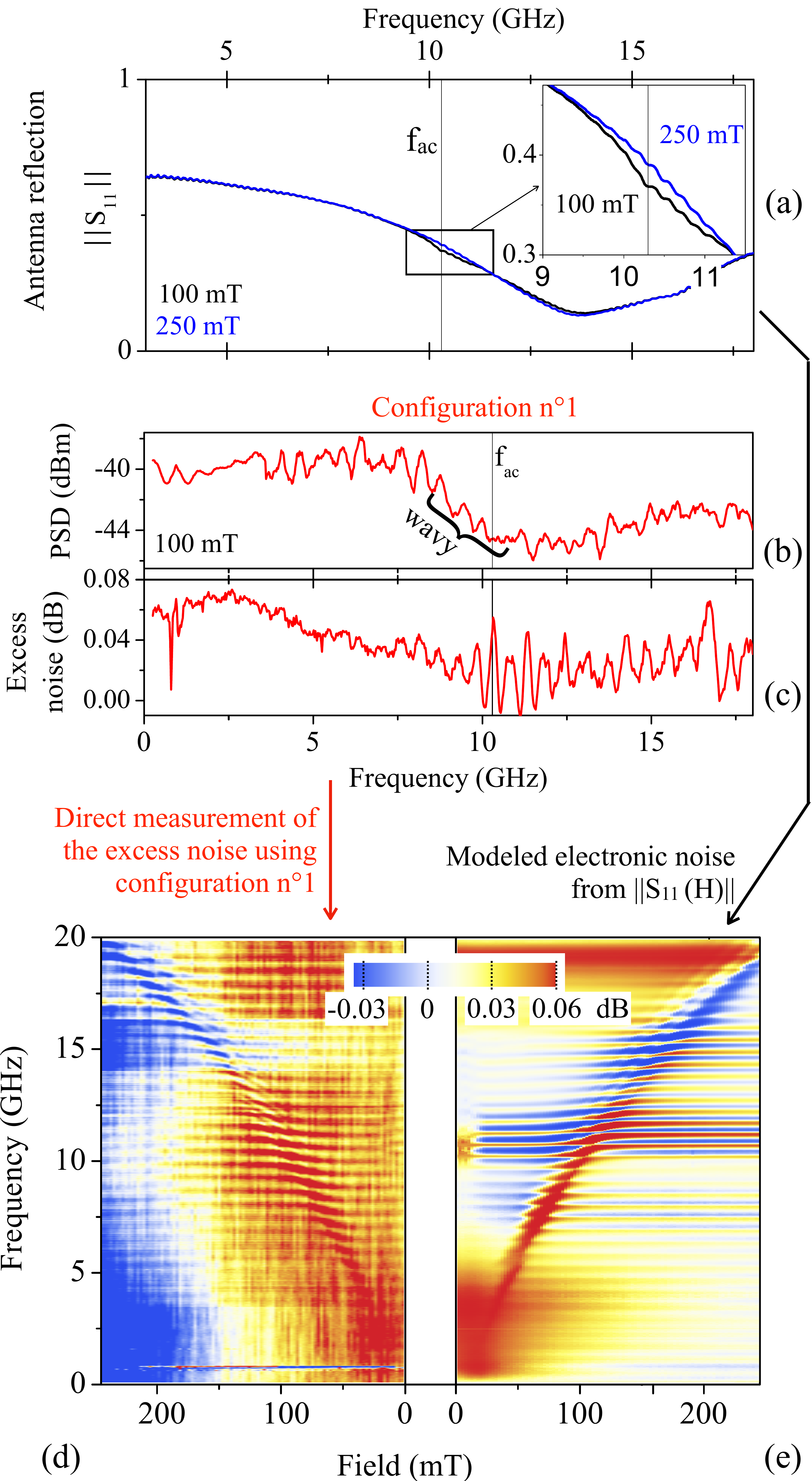}
\caption{Procedure, results and modeling for the first experimental configuration. (a) Reflection coefficient of the antenna for two different applied fields and zoom thereof (inset). (b): Spectral density of the noise power at 100 mT. (c): Excess noise with respect to the $H_\textrm{ref}=0$ reference. The black vertical lines in (a-c) are at a frequency $f_\textrm{ac}$ corresponding to the uniform acoustical mode at 100 mT. The curly brace with the "wavy" label in (b) is meant for comparison with Fig.~\ref{FigureExpProcedureConfig2}(a). (d) Field dependence of the experimental excess noise. The horizontal feature near 1 GHz is an artefact caused by ambient wireless devices. (e) Expected excess Johnson-Nyquist noise of the antenna calculated from Eq.~\ref{PSDelectronic} and the measured field dependence of the antenna impedance. (d) and (e) were decremented with the $H_\textrm{ref}=0$ reference data.}
\label{FigureExpProcedureConfig1}
\end{figure}

\begin{figure*}
\centering
    \includegraphics[width=15 cm]{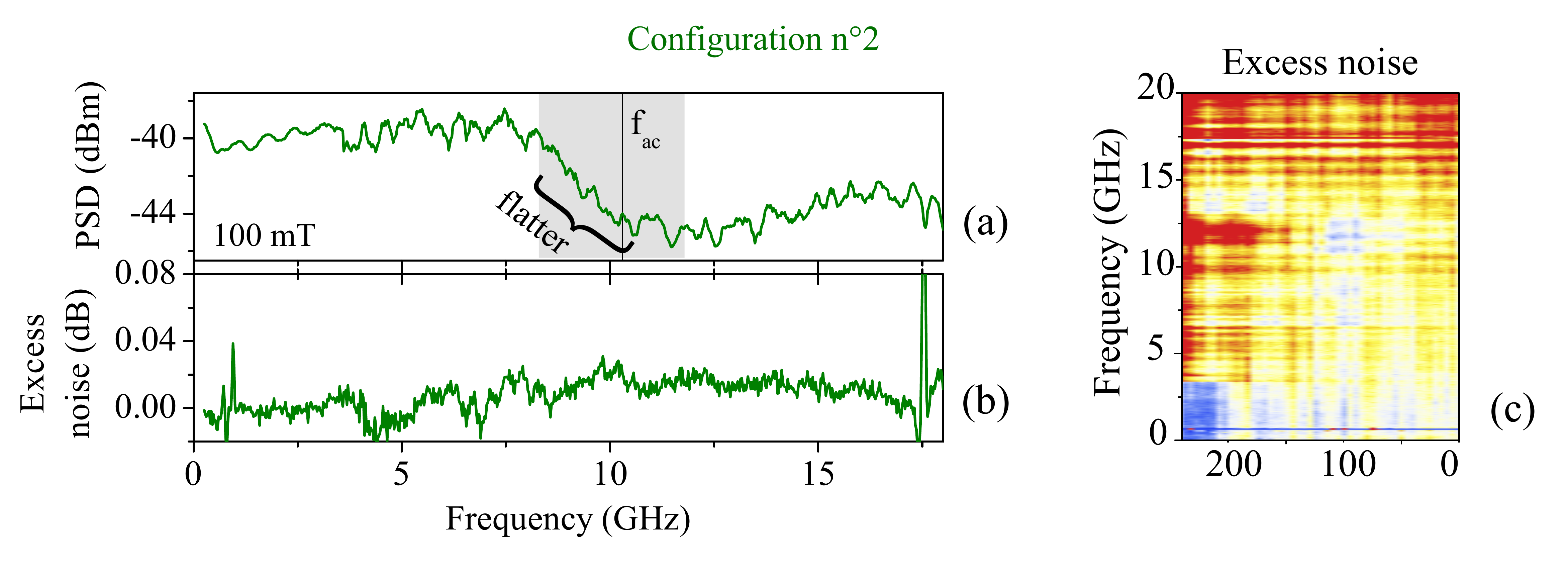}
\hspace*{0.2cm}\caption{Results for the second experimental configuration. (a) Spectral density of the noise
power. The passing band of the isolator appears as the grey zone. (b): Excess noise with respect to the $H_\textrm{ref}=0$ reference. The black vertical line recall the frequency $f_\textrm{ac}$ of the acoustical mode at $k=0$. The curly braces in (a) illustrate the part of the spectrum in which the presence of the isolator strongly attenuates the ripple, in contrast with Fig. 3(b). \textcolor{black}{Panel (c): field dependence of the excess noise, with scales comparable to these of Fig. 4(d).}}
\label{FigureExpProcedureConfig2}
\end{figure*}

The connection between the antenna and the amplifier can be done in a different manner. In the configuration n°1 the microwaves could travel \textit{back and forth} between the antenna and the amplifier's input. In the modified configuration [green in Fig.~\ref{figSetUp}(c)], we insert an isolator before the amplifier; this non-reciprocal element ($||S_{12}^\textrm{iso}|| \leq -20~\textrm{dB}$ while $||S_{21}^\textrm{iso}|| \approx -1~\textrm{dB}$) is meant to ensure that the electromagnetic energy flows \textit{only forward} from the antenna to the remainder of the circuit. The isolator is way better matched than the amplifier and reflects only a tiny part of the electromagnetic energy (here $||S_{11}^\textrm{iso}|| \leq -23~\textrm{dB}$ while the amplifier had $ -17 \textrm{~dB} \leq S_{11}^\textrm{amp} \leq -10~\textrm{dB}$). The goal is to suppress the standing waves within the operation range of the isolator (here 8-12 GHz). As a result, the ripple of the signal in this frequency band is strongly mitigated (compare Fig.~\ref{FigureExpProcedureConfig1}(b) and ~\ref{FigureExpProcedureConfig2}(a) in the range of the "wavy" and "flatter" labels). 
This modified measurement configuration gives strikingly different results. Inserting the isolator seems to fully remove any magnetic information from the field-to-field excess noise, even when the field is chosen such that the SW acoustical frequency is within the passing band of the isolator [compare Fig.~\ref{FigureExpProcedureConfig1}(c) and \ref{FigureExpProcedureConfig2}(b)]. Apart from a non-reproducible drift of the baseline, the spectra of excess noise become featureless for all applied fields, including in the passing band of the isolator (8-12 GHz), \textcolor{black}{see Fig.~\ref{FigureExpProcedureConfig2}(c)}. 

One generally expects the signal arising from the magnetic fluctuations to flow from the antenna towards the amplifier. The apparent suppression of magnetic information in the field-to-field excess noise in the modified experimental configuration may consequently seem counter-intuitive. Let us thus model the antenna noise.

%

%

\section{Magnetic modulation of the Johnson-Nyquist electrical noise power delivered by the measurement circuit} \label{Johnson-Nyquist} 

Let us first evaluate the contribution of the electronic noise to the spectrum recorded by the spectrum analyser in the experimental configuration n°1. 
As already noted, the microwave elements placed between the antenna and the amplifier form a microwave cavity bounded by the impedance mismatch planes $\Pi_1$ and $\Pi_2$ [Fig.~\ref{figSetUp}(c)]. As shown in Fig.~\ref{FigureExpProcedureConfig1}(a), the reflection parameter $\tilde{S}_{11}$ at the antenna boundary of this cavity is field- and frequency-dependent, likely because of the antenna self-inductance. This changes the Johnson-Nyquist noise present in the $\Pi_1-\Pi_2$ cavity. 

This change of the electric noise present in the cavity contains magnetic information; however it has nothing to do with the \textit{amplitude} of the magnetic fluctuations (here: the thermal amplitude). It is related only with the magnetic susceptibility which is included in the inductance of the antenna: this noise contribution scales with the (electronic) temperature $T_e$ of the measurement system. As it is a cavity effect, the fast modulation of the excess noise is present only in the first measurement configuration. Let us estimate the Johnson-Nyquist noise arriving at the amplifier's input. A prerequisite is to evaluate the equivalent impedance presented at the front end of the amplifier. 
\subsection{Impedance shifting within the mismatched cavity} 
Since the antenna is seen through a $\xi \approx 20~\textrm{cm}$-long coaxial cable, the equivalent impedance presented at the front end of the amplifier $\tilde{Z}_\textrm{out}(\xi)$ is distinct from the antenna impedance $\tilde{Z}_\textrm{out}$.  
If the cable has a characteristic impedance $Z_0$ and carries the microwaves with a phase velocity $v_\varphi$, we can shift the reference plane from $\Pi_2$ to $\Pi_1$ and consider that the amplifier is fed by the \{antenna+cable\} system describable as a lumped element of $\xi$-shifted impedance being: 
\begin{equation} \tilde{Z}_\textrm{out}(\xi)=Z_0 \frac{\tilde{Z}_\textrm{out} + i Z_0 \tan(\xi \omega / v_\varphi)}{Z_0 + i \tilde{Z}_\textrm{out} \tan(\xi \omega / v_\varphi)} \label{impedanceShifting} \end{equation}
This \{antenna+cable\} system supplies a Johnson-Nyquist noise of e.m.f. spectral density $4 k_B T_e \mathcal{R}e(\tilde{Z}_\textrm{out}(\xi))$ to the amplifier of input impedance $\tilde{Z}_\textrm{in}(f)$. Assuming a perfect (noiseless) amplifier, the spectral density of the active power delivered to the spectrum analyzer is obtained after passing in the voltage divider configuration, which yields:
\begin{multline}
\textrm{PSD}_\textrm{electronic}(f, \xi, H) [\textrm{in~W/Hz}]=4 k_B T_e ~ \mathcal{R}e\left(\tilde{Z}_\textrm{out}(\xi, H, f)\right)\\ 
~.~ G_\textrm{ampli}(f)  ~.~  \frac{1}{2}  \frac{ \mathcal{R}e (\tilde{Z}_\textrm{in}(f))}{|\tilde{Z}_\textrm{out}(\xi, H,f)+\tilde{Z}_\textrm{in}(f) |^2}
\label{PSDelectronic}
\end{multline} 
where $T_e$ is the temperature of the \{antenna+cable\} system. All terms can be obtained from single-port VNA measurements of the circuit elements.
\subsection{Comparison of the expected electronic noise and the experimental field-to-field excess noise} 
The electronic noise according to Eq.~\ref{PSDelectronic} in excess to the $H_\textrm{ref}=0$ case is evaluated in Fig.~\ref{FigureExpProcedureConfig1}(e). The electronic noise depends on the frequency in an inherently oscillatory manner because of the $\tan(\xi \omega / v_\varphi)$ terms in Eq.~\ref{impedanceShifting}. The plotting of $\textrm{PSD}_\textrm{electronic}(H_\textrm{x}) - \textrm{PSD}_\textrm{electronic} (H_\textrm{ref})$ never succeeds in suppressing this oscillatory contribution, whatever the choice of $H_\textrm{ref}$. Indeed the output impedances of the antenna at reflection coefficient at $H_x$ and at $H_\textrm{ref}$ are unequal, especially near the SW resonances. The circuit-dependent oscillation within the electronic noise must persist near SW resonances whatever field-to-field subtraction is attempted.

The comparison between Figs.~\ref{FigureExpProcedureConfig1}(d) and (e) shows that the main features of the experimental field-to-field excess noise in the conventional measurement configuration are reproduced by the calculated excess electronic noise. The red (positive noise) halo at low fields and low frequencies is reproduced. \textcolor{black}{Also, since the coupling between the microwave photons of the circuit and the magnons below the antenna is included in the experimental value of $\tilde{Z}_\textrm{out}$, the anticrossings between the branches of the microwave photon cavity modes and the branches of the acoustical SWs that were evidenced in the direct noise measurement [Figs.~\ref{FigureExpProcedureConfig1}(d)] are well reproduced in the modeled electronic noise [Figs.~\ref{FigureExpProcedureConfig1}(e)]}. 

From this similarity between the predicted electronic noise and the experimental excess noise, we can infer that the contribution of the magnon noise to the total noise is \textit{at most} a minor part of it. The next section is dedicated to the evaluation of the magnon noise. 
\section{Spectral density of the inductive e.m.f. generated at the antenna by a given population of incoherent spin waves} \label{sectionEmf}%
Let's estimate the variance of the e.m.f. present at the antenna because of the fluctuations of the magnetization in the SAF underneath. The corresponding signal should reach the amplifiers in both experimental configurations. 
We assume that the population of SWs can be described by an effective temperature $T_m$. $T_m$ can be equal to $T_e$ if the spin wave population is at equilibrium like in our present experiments, or it can be larger than $T_e$ if an overpopulation of spin wave is intentionally generated by a way that keeps $T_e$ unchanged. The calculation of the spectral density of this e.m.f. requires to first identify how each SW contributes to the flux picked by the antenna and then to sum over them using the SW density of states.

We will make simplifying assumptions meant to keep an analytical formalism for the description of the spin waves properties, as detailed in the Appendix. We for instance disregard the lateral confinement of the spin waves and assume that they are identical to their counterparts in unbounded SAF films, and that the magnetizations are uniform across the thicknesses of each magnetic film.

\subsection{Relevant inductive signal}%
\begin{figure}
\hspace*{-0.2cm}\includegraphics[width=8 cm]{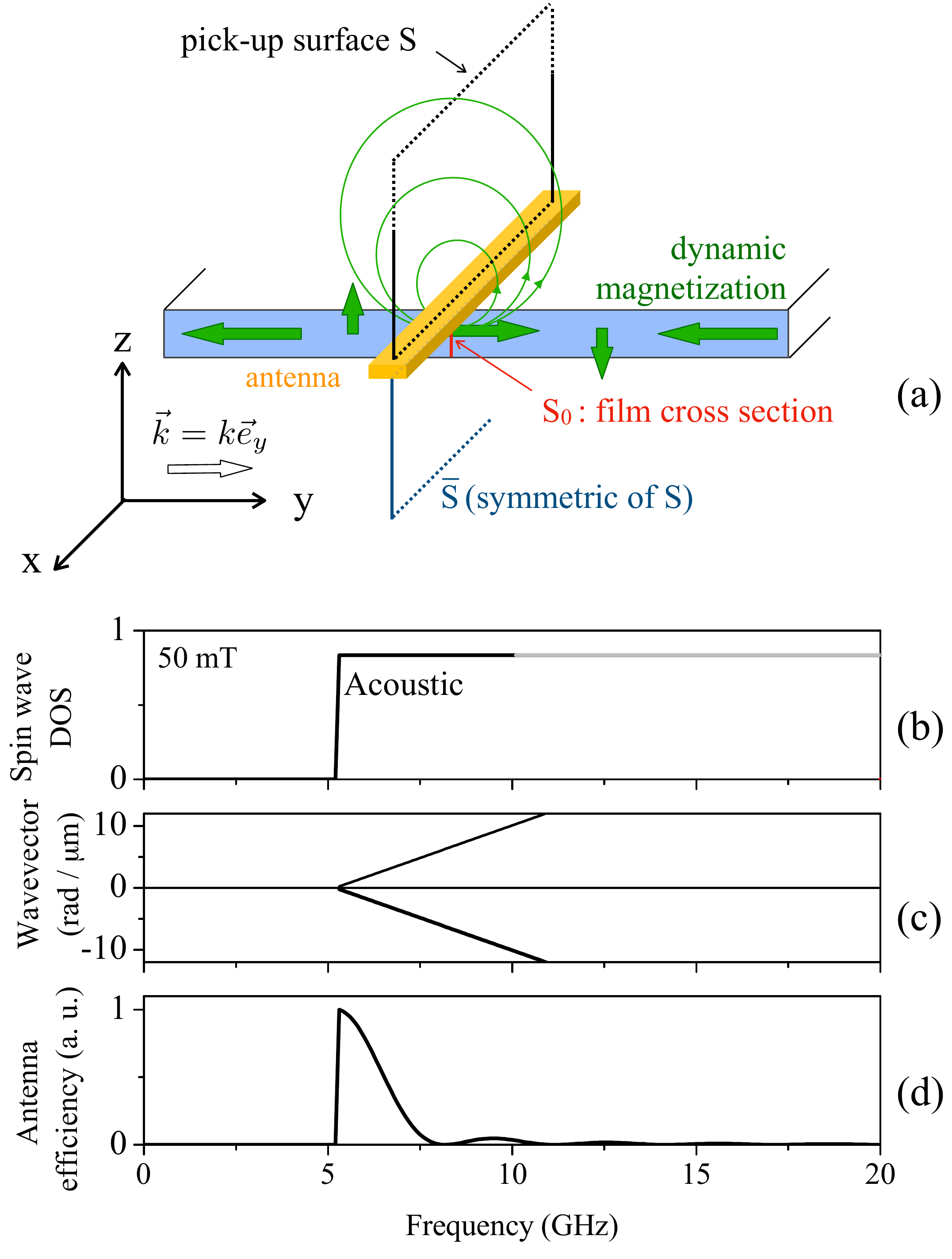}{\centering}
\caption{Principle of the derivation of the power spectral density of the inductive e.m.f. generated by a population of incoherent spin waves (i.e. the magnon noise). (a) Inductive geometry: the antenna collects the flux of the magnetic field lines (green) intercepting the surface $S$ above the magnetic film (blue) of cross section $S_0$, in the presence of a spin wave (green arrows) of wavevector $\vec{k}_y$ perpendicular to the antenna. (b) Sketch of the density of states of the spin waves propagating perpendicularly to the antenna at 50 mT for the acoustic branch. The grey part is when the linear Taylor expansion of the dispersion relation breaks down. (c) Corresponding positive and negative wavevectors. (d) Corresponding antenna efficiency functions $h(k)^2$.}
\label{FigModelisation}
\end{figure}

The inductive e.m.f. is the time derivative of the flux intersecting any surface encompassed by the contour defined by the measurement circuit. The relevant parameters are described in Fig.  \ref{FigModelisation}(a). We can choose the surface $S$, defined to extend vertically from the antenna to $z=+\infty$. We define $\bar{S}$ as the symmetric of $S$, and $S_0$ as the cross section of the SAF stripe. Since the stray field emanating from the magnetic sample decays fast in space, we can assume with marginal error that the antennas are infinitely extended in the $x$ direction. We also assume that the film thickness and film to antenna spacing can be considered as vanishingly small quantities.

Within our assumptions the magnetization precession associated with the spin waves occurs essentially in the $xy$ plane, with a large ellipticity (Table 1); the inductive contribution of the out-of-plane components of the precessing magnetizations is of secondary importance compared to the contribution of the in-plane components, and it shall be neglected. The fluxes $\Phi_{S}$ and $\Phi_{\bar{S}}$ are then equal. Applying Gauss's law ($\vec \nabla .\vec{B}=0$) on the closed surface ${S_0} \cup {S} \cup {\bar{S}}$ [see Fig.~\ref{FigModelisation}(a)], we get that $\Phi_{S} = - \frac{1}{2} \Phi_{S0}$. Applying Lenz's law, the e.m.f. is thus: 
\begin{equation}  \textrm{e.m.f.}_m=  \frac{1}{2} \dot{\Phi}_\textrm{S0}=  \frac{1}{2} \iint_\textrm{S0} \dot{B_y}(x,z) dx dz \label{emf1}
\end{equation}
i.e. it is the time derivative (expressed as the dot symbol) of the half flux of the $B_y$ field in the cross section of the magnet right below the antenna.
%
\subsection{Amplitude of the fluctuation of the total moment for a single spin wave mode}%
The standard deviation of $\Phi_\textrm{S0}$ stems from the standard deviation of the  $y$-component of the total dynamic moment of the two eigenmodes, which is calculated in Annex I. We write the ground state of the two films (labelled 1 and 2) as $\vec M_\textrm{dc}=\{\vec { M}_1^\textrm{dc}, \vec M_2^\textrm{dc} \}$ and the eigenvectors of their $k=0$ spin waves as $\vec M_\textrm{dyn}=\{\vec { M}_1^\textrm{dyn}, \vec M_2^\textrm{dyn} \}$, with the acoustical and optical mode frequencies being $\omega_\textrm{ac}/(2 \pi)$ and $\omega_\textrm{op}/(2 \pi)$ (details in Table 1). We also assume that the population of spin waves can be described by an effective temperature $T_m$. 

As a representative example, the $k=0$ acoustical mode in the scissors state has a fluctuation amplitude for its in-plane component that amounts to:
\begin{equation}    \sqrt{\left(M_1^\textrm{dyn} + M_2^\textrm{dyn}\right)_\textrm{RMS}^2}  =    M_s  \sqrt{\frac{2 k_B T_m}{\mu_0 M_s H_j V}} \label{RMS} ,\end{equation} 
where $V= t \ell_\textrm{mag} w_\textrm{mag}$ is the volume of one of the layers. Eq.~\ref{RMS} follows from the ellipticities of the modes which do not depend much on $k$ when $k t \ll 1$; We shall thus consider that Eq.~\ref{RMS} also applies to all the spin waves considered in our system. 

Summing the inductive contributions over all SWs requires the knowledge of the SW density of states (DOS). Since only the SWs with wavevectors perpendicular to the antenna [red waves in Fig.~\ref{figSetUp}(b)] have a net contribution to the induction flux picked by the antenna, we shall restrict the analysis to the sole longitudinal wavevectors.

\subsection{Density of states within the spin wave manifold} %
In a 1D stripe of length ${\ell_\textrm{mag}}$, the longitudinal wavevectors $k \equiv k_y $ are quantized at values $\frac{2 \pi}{\ell_\textrm{mag}} n$ with $n \in \mathbb N$, and frequency spacings of $ \frac{v_g} {\ell_\textrm{mag}}$.
The SW DOS per unit frequency [Fig.~\ref{FigModelisation}(b)] is the classical result for a 1D system with linear dispersion:
\begin{equation}
\textrm{DOS}(f) =  2 \frac {\ell_\textrm{mag}}{v_g} \mathcal{H} (f - f_\textrm{k=0})~, \label{DOS}
\end{equation}
where $\mathcal{H}$ is the Heaviside function, $v_g = \frac{d \omega}{d|k|}$ is the group velocity (positive here, see Table 1) of the SW branch under study. The factor of 2 recalls the two possible signs of $k_y$.

Note that since the optical branch has a vanishing group velocity for $\vec {k} \perp \vec {H}$ within our approximations (see Table 1), the corresponding DOS is simply a Dirac peak at $f_\textrm{op}$. This cannot be perceived with the RBW settings of the spectrum analyzer. We thus restrict the remainder of this section to the sole acoustical branch.

\begin{table*} 
\caption{Main properties of the spin waves of wavevector $k$ propagating in the direction perpendicular to the applied field in in-plane magnetized isotropic Synthetic AntiFerromagnets. The SAF is assumed to be fully symmetrical and the magnetizations are assumed uniform across the thickness of each of the two magnetic films. The dc field $H_x$ is applied in the $\{1, 0, 0\}$ direction. The dagger symbol recalls that the ground state is unstable against an arbitrary in-plane global rotation when at remanence. The group velocities are defined as the derivative of the angular frequency in the direction of wavevector $\frac {\partial \omega} {\partial |k|} $, i.e. according to the common practice. The interlayer exchange field is defined as $\mu_0 H_{j}=-\frac{2 J}{ {M_{s}}{t}}$. \\}
\label{my table}
\begin{tabular}{| c | c | c | c | c | c |}
 \hline
\textbf{Field}  & $H_x = 0$  & $H_0< H_j$ & $H_0 > H_j$   \\ \hline
\textbf{Ground state}  & $\dagger$ Antiparallel & Scissors & Parallel  \\ 
$M_1^\textrm{dc}/ M_s$  & $\{0, 1 , 0\}$  & $\{ \frac{H_x}{H_j}, \sqrt{1-\frac{H_x^2}{H_j^2}} , 0\}$ & $\{1, 0 , 0\}$  \\ 
$M_2^\textrm{dc}/M_s$   &  $\{0, -1 , 0\}$  & $\{ \frac{H_x}{H_j}, -\sqrt{1-\frac{H_x^2}{H_j^2}} , 0\}$ & $\{1, 0 , 0\}$  \\ \hline
\noalign{\vskip 2mm}

\textbf{Acoustical mode}: \\ \hline
$\omega_\textrm{ac}$ at $k_y=0$ & $\dagger$ 0   & $\gamma_0 H_x \sqrt{\frac{M_s+H_j}{H_j}}$  & $\gamma_0 \sqrt{H_x (H_x+M_s)}$   \\ \hline
$\Delta \omega_\textrm{ac}$ (FWHM) & $\dagger$  $ \alpha \gamma_0 (M_s + H_j) $ & $\alpha \gamma_0 (M_s + \frac{H_j^2+H_x^2}{H_j})$ & $\alpha \gamma_0 (M_s + 2 H_x)$   \\ \hline

Eigenmode: $\vec{M}_1^\textrm{dyn}$ part &  $\dagger$ {$\{-i, 0, 0 \}$}&  {$\left\{-i \sqrt{\frac{M_s+H_j}{H_j}}\frac{\sqrt{H_j^2-H_x^2}}{H_x},+i
   \sqrt{\frac{M_s+H_j}{H_j}}, 1 \right\} $ } & $\left\{0, i\sqrt{\frac{M_s+H_x}{H_x}}, 1 \right\} $ \\ \hline

Eigenmode: $\vec{M}_{2}^\textrm{dyn} $ part & $\dagger$  {$\{+i, 0, 0 \}$} &  {$\left\{+i \sqrt{\frac{M_s+H_j}{H_j}}\frac{\sqrt{H_j^2-H_x^2}}{H_x},+i
   \sqrt{\frac{M_s+H_j}{H_j}}, 1 \right\}$}  & $\left\{0, i\sqrt{\frac{M_s+H_x}{H_x}}, 1 \right\} $   \\ \hline

$\vec{M}_1^\textrm{dyn} + \vec{M}_2^\textrm{dyn} $  & $\dagger$ $\{0, 0, 0\}$ &  {$\left\{0, 2i
   \sqrt{\frac{M_s+H_j}{H_j}}, 2 \right\} $  i.e. $\perp$ to \textit{dc} field} & 2$\left\{0, i\sqrt{\frac{M_s+H_x}{H_x}}, 1 \right\}$   \\ \hline 

$\frac{1}{M_s}(M_1^\textrm{dyn} + M_2^\textrm{dyn})_\textrm{RMS} $   & along y: $ \sqrt{\frac{2 k_B T}{\mu_0 M_s H_j V}}$  &along y: $ \sqrt{\frac{2 k_B T}{\mu_0 M_s H_j V}}$  &  
 along y: $ \sqrt{\frac{2 k_B T}{\mu_0 M_s H_x V}}$   \\ \hline

$v_g^\textrm{ac}$, $\vec{H}_x \perp \vec{k}_y$, 
$k>0$ &  0 & 
 + $\frac{1}{2} \gamma_0 M_s t$ $\frac{H_0}{H_j} \frac{M_s}{\sqrt{H_j(H_j+M_s)}}$  &  + $\frac{1}{2} \gamma_0 M_s t$ ${\frac{M_s}{\sqrt{H_x(H_x+M_s)}}}$  \\  \hline

   
\noalign{\vskip 2mm}

\textbf{Optical mode}: \\ \hline
$\omega_\textrm{op}$  at $k=0$ & $\gamma_0 \sqrt{M_s H_j}$  & $\gamma_0 \sqrt{\frac{M_s}{H_j}} \sqrt{H_j^2-H_x^2}$ &           $ \gamma_0 \sqrt{(H_x-H_j) (H_x+M_s-H_j)}$   \\ \hline
$\Delta\omega_\textrm{op}$ (FWHM) & $ \alpha \gamma_0 (M_s + H_j) $  &  $\alpha \gamma_0 (M_s + \frac{H_j^2-H_x^2}{H_j})$& $\alpha \gamma_0 (M_s + 2 H_x - 2 H_j)$   \\ \hline

$M_1^\textrm{op} $ eigenstate&    {$\left\{ +i \sqrt{\frac{M_s}{H_j}},  0, -1 \right\}$}   &  {$\left\{+i \sqrt{\frac{M_s}{H_j}},  -i \sqrt{\frac{M_s}{H_j}} \frac{H_x    }{\sqrt{H_j^2-H_x^2}}, -1 \right\} $} 
 &  {$\{0, \sqrt{\frac{H_x+M_s-H_j}{H_x-H_j}}, -1 \} $}   \\ \hline

$M_2^\textrm{op} $ eigenstate&    {$\left\{ +i \sqrt{\frac{M_s}{H_j}},  0, 1 \right\}$}   &   {$\left\{
+i \sqrt{\frac{M_s}{H_j}},  + i \sqrt{\frac{M_s}{H_j}} \frac{H_x    }{\sqrt{H_j^2-H_x^2}}, +1 \right\}$}  
 &  {$\{0, -\sqrt{\frac{H_x+M_s-H_j}{H_x-H_j}}, 1 \}$}   \\ \hline

$M_1^\textrm{op} + M_2^\textrm{op}$ eigenstate &  {$\{2i \sqrt{\frac{M_s}{H_j}}, 0, 0\}$}  &    {$\{2i \sqrt{\frac{M_s}{H_j}}, 0, 0\}$ i.e. $\parallel$ to \textit{dc} field} & $\{0,0, 0\}$   \\ \hline

$\frac{1}{M_s}(M_1^\textrm{op} + M_2^\textrm{op})_\textrm{RMS} $  &    along x: $\sqrt{\frac{2 k_B T}{\mu_0 M_s H_j V}} $ & along x: $\sqrt{\frac{2 k_B T}{\mu_0 M_s H_j V}} $   & along x: 0     \\ 

 \hline
 
$v_g^\textrm{op}$, $\vec{H_x} \perp \vec{k_y}$, 
$k>0$ & 0  & 
0&    0  \\  \hline

\noalign{\vskip 2mm}
\end{tabular}
\end{table*}

\subsection{Power spectral density of the e.m.f. due to magnetic fluctuations} 
To finally get the power spectrum of the $\textrm{e.m.f.}_m$, we still need to account for the variable part of the flux picked-up by the antenna. This is done by writing:
\begin{equation}\frac{ d B_y}{dt} = i \omega B_y^\textrm{dyn} \label{ByversusMy} \end{equation}
To first order in $k$, the difference between the flux density and the dynamic magnetization \textit{within} the magnetic sample is simply $B_y^\textrm{dyn} = \mu_0 M_y^\textrm{dyn} (1 - \frac{k t}{2})$.
This flux is partly transduced to an antenna voltage with the dimensionless antenna efficiency function \cite{ sushruth_electrical_2020} that only depends on the antenna width $\mathcal{A}$ and spacing between the mid-plane of the antenna and the mid-plane of the magnetic film $s_\textrm{eff} \approx 250~\textrm{nm}$:
\begin{equation} h (k)=\frac{\sin(k \mathcal{A}/2)}{k \mathcal{A}/2} e^{-|k|s_\textrm{eff}} \label{Aefficiency} \end{equation}
Using Eq.~\ref{emf1}, \ref{DOS}, \ref{ByversusMy},  this yields the spectral density of the inductive 'noise' delivered by the population of incoherent magnons for a given branch: 
\begin{multline} 
\frac{d}{df} \textrm{e.m.f.}_m^2 ~[\textrm{in~units~of~}\textrm{V}^2/\textrm{Hz}]= \\ \frac{ 2 \omega^2}{v_g} t w_\textrm{mag} \mu_0^2 \left[V \left(M_{1y}^\textrm{dyn}+M_{2y}^\textrm{dyn} \right)_\textrm{RMS}^2 \right]  h (k)^2  \left(1 - \frac{k t}{2} \right)^2
\label{emf2}
\end{multline}
Note that the term enclosed between square brackets is in fact independent from the sample geometry and volume $V$, as can be seen from Eq.~\ref{RMS}. Note also that $k$ is implicitly defined from the frequency by the dispersion relation, i.e. to first order: $k= \frac{1}{v_g} (\omega-\omega_{k=0})$ [Fig.~\ref{FigModelisation}(c)]. 

This e.m.f. is delivered in the $\Pi_2$ plane where the amplifier is seen as a lumped element of $\xi$-shifted input impedance $\tilde{Z}_\textrm{in}(\xi, f)$ which is:
\begin{equation}
\tilde{Z}_\textrm{in}(\xi, f)=Z_0 \frac{\tilde{Z}_\textrm{in} + i Z_0 \tan(\xi \omega / v_\varphi)}{Z_0 + i \tilde{Z}_\textrm{in} \tan(\xi \omega / v_\varphi)}
\end{equation}
such that the active power delivered to the spectrum analyser by the acoustical branch of the spin waves when in the scissors state is:
\begin{multline}
\textrm{PSD}_\textrm{mag.fluct.}(f, \xi, H) [\textrm{in~W/Hz}]=4 k_B T_m ~ 2\mu_0 w_\textrm{mag}\omega_\textrm{ac}  \\
~\mathcal{H}(f-f_\textrm{ac}) 
\times \left[h (k)^2  \left(1 - \frac{k t}{2} \right)^2 \right] \frac{M_s+H_j}{M_s} \\
~.~ G_\textrm{ampli}(f)  ~.~  \frac{1}{2}  \frac{ \mathcal{R}e (\tilde{Z}_\textrm{in}(\xi, f))}{|\tilde{Z}_\textrm{in}(\xi,f)+\tilde{Z}_\textrm{out}(f, H) |^2}
\label{emfacou}
\end{multline}
%
\subsection{Color and amplitude of the inductive magnon noise}

Several point are worth noticing in Eq.~\ref{emfacou}.

(i) The inductive magnetic noise is colored and narrow-band, essentially thanks to the finite k-vector efficiency of the antenna: the noise should suddenly shoot up at the acoustical frequency and then extend up to the first zero of the $h(k)$ function occurring at $f_\textrm{ac} + \frac{v_g}{\mathcal{A}}$. This contrasts with the focused BLS configuration for which the group velocity in all directions has to be taken into account and the antenna width $\mathcal{A}$ has to be replaced by the laser spot size.

(ii) Neither $M_s$ nor $t$ appear explicitly in the inductive magnetic noise (Eqs.~\ref{emfacou}). This could seem in contradiction with the intuition that an inductive signal should scale with the total magnetic moment, as it does for each value of $k$. Lifting this apparent contradiction requires to notice that we have also $v_g \propto t$: the signal for each $k$ value is spread over a proportionally large frequency interval, yielding a plateau of $\textrm{PSD}_\textrm{mag.fluct.}$ that is independent from the magnetic thickness $t$ for $f \approx f_\textrm{ac}$. However the frequency interval in which the magnon noise is substantial is typically $f_\textrm{max}-f_\textrm{ac} = \frac{v_g^\textrm{ac}}{\mathcal{A}}$ which scales proportionally to $ \gamma_0 M_s t$ (see Table 1): in brief, the plateau of the magnon noise is independent from $t$ but its integral is proportional to $M_s t $, in line with the qualitative expectation.

(iii) If the amplifier was hypothetically perfectly matched with $\tilde{Z}_\textrm{in}=Z_0$ (or equivalently if a perfectly matched isolator is placed at the front end of the amplifier), then the antenna would simply see the lumped impedance $\tilde{Z}_\textrm{in}(\xi)=Z_0$. The $\textrm{PSD}_\textrm{mag.fluct.}$ would not oscillate with the frequency; the color of this noise would reflect the product of the SW DOS and the antenna efficiency function. In the realistic case when the amplifier is not perfectly impedance matched (i.e. configuration n°1), $\textrm{PSD}_\textrm{mag.fluct.}$ exhibits a ripple along a frequency comb of spacing $v_\varphi / \xi$, i.e. in a manner very similar to the electronic noise so that they can't be easily distinguished. The most important question is thus the respective amplitudes of the magnetic noise (Eq.~\ref{emfacou}) and the formerly modeled electronic noise (Eq.~\ref{PSDelectronic}).

\section{Discussion on the inductive measurement of spin wave populations}
\subsection{Respective contributions of the spin wave population-induced noise and the electronic noise at thermal equilibrium}
It is primordial to notice that the term $ 2\mu_0 w_\textrm{mag}$ in Eq.~\ref{emfacou} has the dimension of an \textit{inductance}, so the term $2\mu_0 w_\textrm{mag}\omega_\textrm{ac} $ has the dimension of an impedance. For a stripe width $w_\textrm{mag} =20~\mu\textrm{m}$, this so-called "magnon" impedance amounts typically to $0.6-6~\Omega$ in the 2-20 GHz interval considered here.
This impedance should be compared with the much larger impedance of the other noise sources present in the measurement chain (Eq.~\ref{PSDelectronic}) with our initially chosen geometry. 

This comparison of impedances evidences that if the circuit temperature $T_e$ and that of the spin wave bath $T_m$ are equal (like in our experiments) the noise power inductively supplied by the spin waves in a $w_\textrm{mag} =20~\mu\textrm{m}$ stripe cannot be expected to be more than a tiny fraction of the total noise supplied by the circuit. This explains the finding that the modified experimental configuration collects much less magnetic signal than the conventional experimental configuration.

Rendering the magnon noise perceivable within the total noise would require Eq.~\ref{emfacou} to exceed Eq.~\ref{PSDelectronic}, which occurs when the magnon impedance of the antenna exceeds the input impedance of the amplifier, i.e. typically when:
\begin{equation}T_m > T_e \frac{50 ~\Omega}{2 \mu_0 w_\textrm{mag} \omega_\textrm{ac} } \label{perceivableCriterion}\end{equation}
If the spin wave bath is at equilibrium with the circuit (i.e. if $T_m = T_e$) then the previous condition is not met in our system. One could obviously increase the width $w_\textrm{mag}$ of the magnetic stripe to collect magnon noise from a longer region; meanwhile, it would be beneficial to adjust the length and thickness of the antennas to get a better matching between $\tilde{Z}_\textrm{out}$ and $Z_0$ and thereby to minimize the frequency ripple of the electronic noise that is difficult to correct.
However the rigorous subtraction of the electronic noise to get the sole magnon noise would still be impossible from the sole field-dependence of the noise delivered by the antenna.
\subsection{Quantifying spin wave populations using a two-magnetic-temperature measurement}
There is fortunately another possibility/situation to quantify inductively the population of spin waves, notably when the latter is \textit{not} in equilibrium with the circuit. Such a situation is for instance encountered in a spin-torque oscillator, and one could think of coupling the latter to an inductive antenna. When a spin-torque current $I_\textrm{STT} $ is applied in a sub-threshold (non-oscillating) regime, overpopulations of incoherent spin waves build up far beyond the thermal expectation values \cite{petit_spin-torque_2007} within the free layer of the oscillator. The metric $\textrm{PSD}(I_\textrm{STT} \neq 0) - \textrm{PSD}(I_\textrm{STT} = 0)$ can be used to estimate the spin wave population $T_m$, although the common practice is to express this metric as a precession angle after having assumed that a single spin wave mode exists in the system.
Another equally relevant situation relies in various non-linear FMR experiments, when the pumped coherent populations of spin waves coexist with incoherent populations resulting from multimagnon processes. This happens for instance in the transient route towards equilibrium when cascades of four-magnon scattering events have partitioned the excess energy initially injected  by parametric pumping to a single pair of modes \cite{schultheiss_direct_2012}. 

The method to count a spin wave overpopulation can be directly inspired from the 2-temperature methods implemented to measure the noise figures of rf receivers: Instead of performing a field-to-field subtraction of some reference noise spectrum, one can perform a \textit{hot-to-cold} comparison of the noise spectra. Indeed, since the electronic noise (Eq.~\ref{PSDelectronic}) is independent from the magnon noise (Eq.~\ref{emfacou}), it can be subtracted if two measurements are performed at the \textit{same} circuit temperature $T_e$ and the \textit{same} matching conditions (hence \textit{same} applied field) but with two \textit{different} levels of spin populations that we shall describe with their effective temperatures $T_{m1}$ and $T_{m2}$. 

One of the temperatures must be known (for instance the reference measurement is $T_{m1}=T_e=300~\textrm{K}$). One needs also to assume that the spin wave populations do not affect their resonance frequencies (i.e. neither blue nor red shift). In this case one can use Eq.~\ref{emfacou} to deduce the spin wave overpopulation $(T_{m2}-T_{m1})$ in a rigorous manner. Once again, adjusting the electrical parameters of the antenna to get the best achievable matching between $\tilde{Z}_\textrm{out}$, $Z_0$ and $\tilde{Z}_\textrm{in}$ and thereby minimizing the frequency ripple can only help to minimize the errors when evaluating the electrical terms within Eq.~\ref{emfacou}.

\section{Summary and concluding remarks}
We have studied the noise present in the microwave signal collected by an inductive antenna on top of a thin magnetic stripe and its spin wave bath. The spin waves are --by definition-- (inductively) coupled to the microwave photons of the antenna, such that one may question whether it is possible to measure the populations of spin waves in a non-invasive manner while collecting the microwave photons of the circuit. 

We have found that the experimental noise power delivered by the antenna to the measurement circuit is clearly correlated with the spin wave frequencies. Our modeling indicates that this noise power contains two contributions: the magnon noise, i.e. the contribution of the population of spin waves coupled to a noise-free measurement circuit (Eq.~\ref{emfacou}), and that of the electronic thermal fluctuations coupled to a noise-free (but with a finite susceptibility) magnetic body (Eq.~\ref{PSDelectronic}). This Johnson-Nyquist noise is modulated by the antenna reflection coefficient in a way that depends on the magnetic susceptibility and on the electrical parameters of the microwave cavity formed by the measurement circuit if not perfectly impedance-matched.

Since the electronic noise also contains magnetic information, the two sources of noise cannot be separated by solely field-dependent measurements. Measurements performed at different levels of spin wave populations, i.e. at different spin wave effective temperatures, can circumvent this difficulty. This opens the way for the inductive quantitative measurement of spin wave populations that are out-of-equilibrium in nanoscale systems. This could be applied to a great advantage to the situations when the spin wave overpopulations result from non-linear processes like for instance parametric pumping \cite{schultheiss_excitation_2019}.

The authors acknowledge financial support from the French National Research Agency (ANR) under Contract No. ANR-20-CE24-0025 (MAXSAW). M.S.N. and J.L. acknowledge in addition the FETOPEN-01-2016-2017 [FET-Open research and innovation actions (CHIRON project: Grant Agreement ID: 801055)]. We thank C. Dupuis, D. Bouville, A. Harouri and N. Bardou for technical assistance for the sample fabrication.

\section*{Appendix: analytics of the spin waves in synthetic antiferromagnetics}
This appendix gives further details on analytical description of the SW properties in SAFs. The algebra is done for a SAF but the results listed in Table I can be easily transposed to a single layer film magnetized along the in-plane applied field. One should simply consider that the interlayer exchange coupling field $H_j$ vanishes while not forgetting that for a correct accounting of the dipole-dipole interaction, the single layer film must be described with a thickness equal to $2t$.
\subsection{Simplifying assumptions and methods}
We assume that the SAF can be described by two films (labelled 1 and 2) that are both uniformly magnetized across their thickness. 
Keeping the same notations as in the main part of this paper, we write the ground state as $\vec M_\textrm{dc}=\{\vec { M}_1^\textrm{dc}, \vec M_2^\textrm{dc} \}$. For an applied field $0< H_x < H_j$, the SAF is in the scissors state \cite{stamps_spin_1994,  worledge_spin_2004, devolder_spin_2012}. The dynamical matrix of the system can be calculated following the standard methodology \cite{grunberg_magnetostatic_1981, nortemann_microscopic_1993, zhang_angular_1994, grimsditch_magnetic_2004, giovannini_spin_2004, henry_propagating_2016}. The real and imaginary parts of its eigenvalues are the frequencies of the SWs and their half linewidths. The complex-valued eigenvectors $\vec{M}_\textrm{dyn}$ are the dynamic magnetization components $\vec M_\textrm{dyn}=\{\vec { M}_1^\textrm{dyn}, \vec M_2^\textrm{dyn} \}$, with the dephasing of the components being coded in their complex argument. 

The results of the eigenvalues and eigenvectors calculations are summarized in Table 1 and illustrated in Fig.~\ref{FigModelisation}(b). The first mode is 'acoustical' \cite{stamps_spin_1994}: the two magnetizations precess while keeping constant the angle between their in-plane projections [when seen in top view, this is a rigid rocking of the scissor state at a frequency $\omega_\textrm{ac}/(2 \pi)$]. The second mode is 'optical': the scissoring angle breathes at a frequency $\omega_\textrm{op}/(2 \pi)$. 

The group velocity of the acoustical and the optical branches are gathered in Table 1. The corresponding DOS and dispersion relations are illustrated in Fig.~\ref{FigModelisation}(b, c). The group velocities were found with the procedure of ref.~\cite{nortemann_microscopic_1993} while augmenting the dynamical matrix with the contribution of finite wavevectors to the self- and mutual demagnetizing effects of the two magnetic layers according to Eq. 25 and 26 of ref.~[\onlinecite{henry_propagating_2016}]. The eigenvalues of the augmented dynamical matrix were Taylor-expanded to first order in $k t$ at $k=0$ to get the $v_g$'s of the acoustical and optical spin wave branches in the small $k$ limit. In our specific case of $\vec{k}_y \perp \vec{H_x}$, the dispersion relations of the acoustical and optical spin waves are both reciprocal.
\subsection{Amplitude of the magnetization fluctuations}
The eigenvectors $\vec {M}_1^\textrm{dyn}$ and $\vec M_2^\textrm{dyn}$ at $k=0$ can be used to evaluate the amplitude of the magnetic fluctuations in the uniform modes. For both acoustical and optical modes, the dynamical component $\vec {M}_z^\textrm{dyn}$ is in quadrature with the in-plane component $\vec{M}_{xy}^\textrm{dyn}$: the magnetizations recover an in-plane orientation in a synchronous manner twice per period.
Noticing that $k_B T_m \gg \hbar \omega_\textrm{ac},~ \hbar \omega_\textrm{opt}$, we can use the  equipartition theorem to link the amplitude of magnetic fluctuations with the effective temperature  $T_m$ of the magnon bath. The standard deviations of the in-plane magnetization fluctuations related to a given eigenmode follow from the average energy of that degree of freedom, i.e.:
\begin{equation} E(\vec{M}_\textrm{dc}+ \epsilon \vec{M}^{xy}_\textrm{dyn})-E(\vec{M}_\textrm{dc}) = \frac{1}{2} k_B T_m~,\end{equation}
where $\epsilon$ is a dimensionless number whose square is proportional to the number of magnons of the considered type. Solving for $\epsilon \ll 1$ to second order in $\epsilon$ yields the in-plane amplitude of the fluctuations. The derivation of the e.m.f. (Eq.~\ref{emf1}) requires to vector-sum over the two layers. Table 1 gathers the $\frac{1}{M_s}(M_1^\textrm{dyn} + M_2^\textrm{dyn})_\textrm{RMS} $, i.e. the standard deviations of the $y$-component of this vector sum for the two eigenmodes. 
\subsection{Validity of the expressions listed in Table I}
The results listed in the last column of Table 1 would strictly hold only in the ultrathin limit at $t=0$. At finite applied fields, vertical gradients of the magnetization orientation develop within the magnetic layers unless their thickness $t$ is much smaller than the two characteristic lengths $A/J$ and $\sqrt{2A_\textrm{ex}/(\mu_0 M^2_s)}$, where $A_\textrm{ex}$ is the exchange stiffness. In our samples, we estimate these two lengths to be respectively 12 nm and 4 nm, i.e. thinner than the $t=17$ nm of our two CoFeB layers. In this situation, the magnetizations of the magnetic regions directly touching the Ru spacer stay more antiparallel that the other regions, such the 2-macrospin description of the ground state gets inaccurate at finite applied fields. Micromagnetic calculations (not shown) lead us to conclude that the acoustical mode, which involves less the interlayer exchange interaction than the optical mode, is still reasonnably quantitatively described by the expressions listed Table I. This is not the case of the optical mode, especially when $H_x$ is comparable to $H_j$. A collateral consequence is that the characteristic field $H_{j}$ does not correspond to the saturation of the magnetization in any part the stack, the saturation being gradual within the thickness of the films. For our specific film thickness, the expressions for the optical mode listed in Table I should thus be considered as qualitative only.

%

\end{document}